\documentclass[aip,apl,twocolumn,preprintnumbers,amsmath,amssymb]{revtex4}

\usepackage{graphicx}
\usepackage[colorlinks=true,citecolor=blue]{hyperref}

\begin{document}
    \title{Cavity Quantum Electrodynamics with a Single Quantum Dot Coupled to a Photonic Molecule}
    \author{Arka Majumdar}
    \email{arkam@stanford.edu}
    \author{Armand Rundquist}
    \author{Michal Bajcsy}
    \author{Jelena Vu\v{c}kovi\'{c}}
    \affiliation{E.L.Ginzton Laboratory, Stanford University, Stanford, CA, $94305$\\}
\begin{abstract}
We demonstrate the effects of cavity quantum electrodynamics for a
quantum dot coupled to a photonic molecule, consisting of a pair
of coupled photonic crystal cavities. We show anti-crossing
between the quantum dot and the two super-modes of the photonic
molecule, signifying achievement of the strong coupling regime.
From the anti-crossing data, we estimate the contributions of both
mode-coupling and intrinsic detuning to the total detuning between
the super-modes. Finally, we also show signatures of off-resonant
cavity-cavity interaction in the photonic molecule.
\end{abstract}
\maketitle
A single quantum dot (QD) coupled to a photonic crystal (PC)
cavity is an important building block for integrated nanophotonic
quantum information processing devices \cite{andrei_njp}. This
solid-state cavity quantum electrodynamic (cQED) system is of
considerable interest to the quantum optics community for the
generation of non-classical states of light
\cite{AF_natphys,AM_tunneling}, for its application to all-optical
\cite{arka_switching,edo_switching} and electro-optical switching
\cite{AF_eom}, and due to unusual effects like the off-resonant
dot-cavity interaction due to electron-phonon coupling
\cite{article:majumdar09}. However, all of the cQED effects
demonstrated so far in this system involve a single cavity.
Although numerous theoretical proposals employing multiple
cavities coupled to quantum dots exist in the literature
\cite{cca_plenio,cca_andrew,ciuti_fermionized_photon},
experimental development in this direction is rather limited.
Recently it has been reported that strongly sub-Poissonian light
can be generated from a pair of coupled cavities containing a
single QD \cite{savona_ph_molecule, imma_ph_molecule}. This double
cavity, also called a photonic molecule, coupled to a single QD
forms the first step towards building an integrated cavity network
with coupled QDs. Photonic molecules made of PC cavities were
studied previously \cite{kapon_2009, kapon_2011} to observe
mode-splitting due to coupling between the cavities. In those
studies, a high density of QDs was used merely as an internal
light source to generate photoluminescence (PL) under above-band
excitation and no quantum properties of the system were studied.
In another experiment, a photonic molecule consisting of two
micropost cavities was used along with a single QD to generate
entangled photons via exciton-biexciton decay, but the QD-cavity
system was in the weak coupling regime and the Purcell enhancement
was the only cQED effect observed \cite{snellart_ph_mol}.

In this paper, we demonstrate strong coupling of a photonic
molecule with a single QD. We show clear anti-crossing between the
QD and two super-modes formed in the photonic molecule. In
general, the exact coupling strength between two cavities in a
photonic molecule is difficult to calculate, as the observed
separation between the two modes has contributions both from the
cavity coupling strength as well as from the mismatch between the
two cavities due to fabrication imperfections. However, by
monitoring the interaction between a single QD and the photonic
molecule we can exactly calculate the coupling strength between
the cavities and separate the contribution of the bare detuning
due to cavity mismatch. In fact, without any coupling between two
cavities, one cannot have strong coupling of the QD with both of
the observed modes. Hence, the observed anti-crossing of the QD
with both modes clearly indicates coupling between the cavities.
Apart from the strong coupling, we also demonstrate off-resonant
phonon-mediated interaction between the two cavity modes, a
recently found effect in solid-state cavity systems.

\begin{figure}
\centering
\includegraphics[width=3.25in]{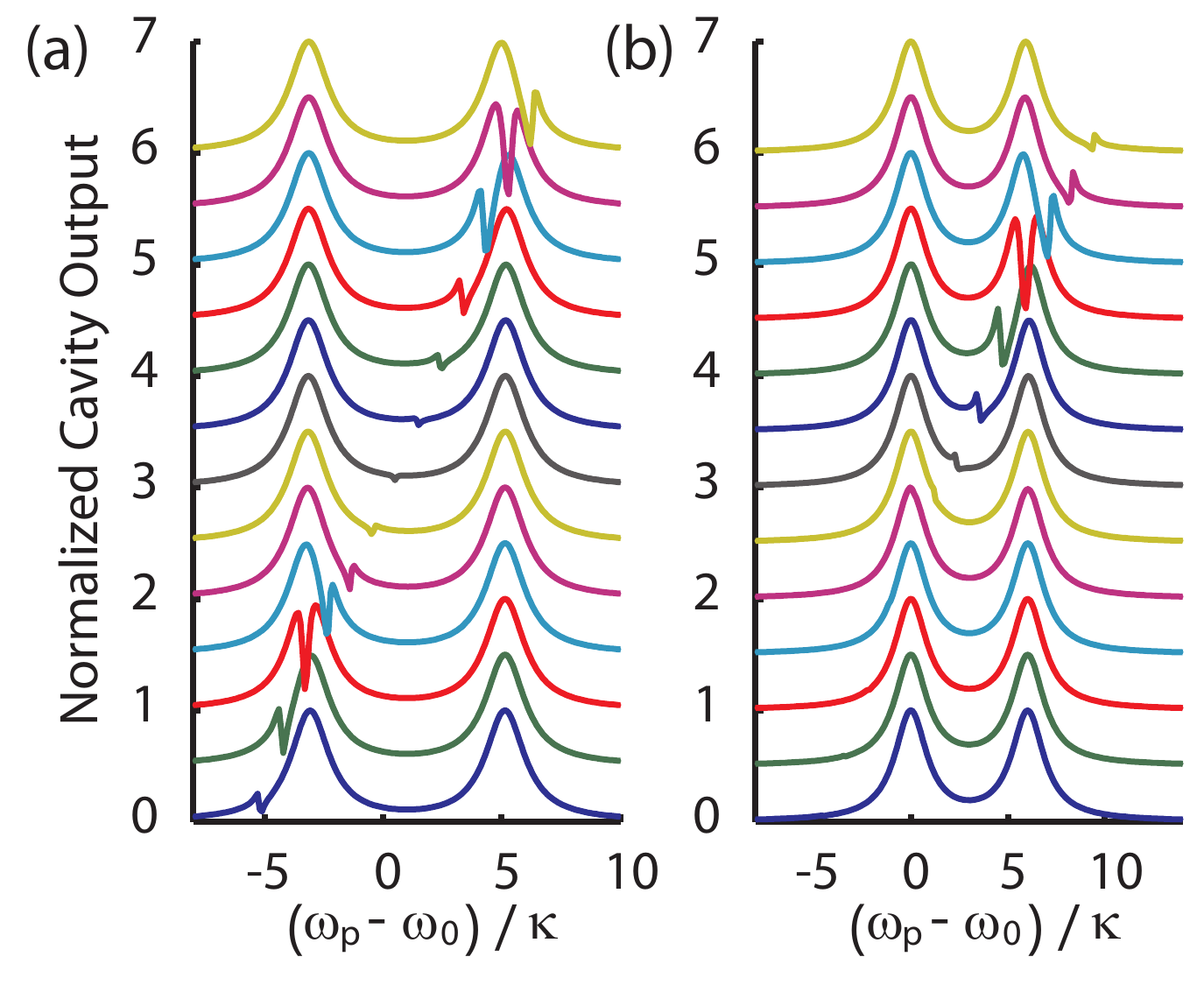}
\caption{(color online) Numerically calculated cavity transmission
spectra when the QD resonance is tuned across the two cavity
resonances. (a) Anticrossing is observed between the quantum dot
and both cavity modes when the two cavities are coupled (coupling
rate between the two cavities is $J/2\pi=80$ GHz). (b) When the
two cavities are not coupled ($J=0$), we observe anti-crossing in
only one cavity. Parameters used for the simulation: cavity decay
rate $\kappa/2\pi=20$ GHz (for both cavities); QD dipole decay
rate $\gamma/2\pi$=1 GHz; dot-cavity coupling rate of $g/2\pi=10$
GHz; intrinsic detuning between the bare cavity modes
$\Delta_o/2\pi=40$ GHz for (a) and $120$ GHz for (b). The plots
are vertically offset for clarity. The horizontal axis corresponds
to the detuning of the probe laser frequency $\omega_p$ from the
cavity $a$ resonance $\omega_0$ in units of cavity field decay
rate.} \label{phot_mol_theory}
\end{figure}

Let us consider a photonic molecule consisting of two cavities
with annihilation operators for their bare (uncoupled) modes
denoted by $a$ and $b$, respectively.  We assume that a QD is
placed in and resonantly coupled to the cavity described by
operator $a$. The Hamiltonian describing such a system is:
\begin{equation}
\mathcal{H} =\Delta_o b^\dag b+J(a^\dag b+a b^\dag)+g(a^\dag
\sigma +a\sigma^\dag)
\end{equation}
where $\Delta_o$ is the detuning between the two bare cavity
modes; $J$ and $g$ are, respectively, the inter-cavity and
dot-cavity coupling strength; $\sigma$ is the QD lowering
operator; and the resonance frequency $\omega_0$ of the cavity
with annihilation operator $a$ is assumed to be zero. We now
transform this Hamiltonian by mapping the cavity modes $a$ and $b$
to the bosonic modes $\alpha$ and $\beta$ introduced as
$a=\cos(\theta) \alpha+\sin (\theta) \beta$ and $b=\sin(\theta)
\alpha-\cos(\theta) \beta$. We note that this mapping maintains
the appropriate commutation relations between operators $a$ and
$b$. Under these transformations we can decouple the two cavity
modes ($\alpha$ and $\beta$) for the following choice of $\theta$:
\begin{equation}
\label{cond} \tan (2\theta)=-\frac{2J}{\Delta_o}
\end{equation}
Under this condition the transformed Hamiltonian becomes:
\begin{eqnarray*}
\mathcal{H}&=&\alpha^\dag \alpha (\Delta_o\sin^2(\theta)+J
\sin(2\theta))+g \cos(\theta)(\alpha^\dag \sigma+\alpha\sigma^\dag)\\
 &+&\beta^\dag \beta(\Delta_o\cos^2(\theta)-J \sin(2\theta))+g
\sin(\theta)(\beta^\dag \sigma+\beta\sigma^\dag)
\end{eqnarray*}
Therefore, a QD coupled to a photonic molecule has exactly the
same eigen-structure as two detuned cavities with the QD coupled
to both of them (from the equivalence of the two expressions above
for the Hamiltonian $\mathcal{H}$). The super-modes of the
transformed Hamiltonian $\alpha$ and $\beta$  will be separated by
$\Delta=\sqrt{\Delta_o^2+4J^2}$ (obtained by subtracting the terms
multiplying $\alpha^\dag \alpha$ and $\beta^\dag \beta$, under the
conditions of Eq.\ref{cond}) and the interaction strength between
the QD and the super-modes will be $g_1=g \cos(\theta)$ and $g_2=g
\sin(\theta)$. If the two cavities are not coupled ($J=0$ and
$\theta=0$), we can still observe two different cavity modes in
the experiment due to $\Delta_o$, the intrinsic detuning between
two bare cavities. However, if we tune the QD across the two
cavities in this case, we will observe QD-cavity interaction only
with one cavity mode (in this case $\alpha$, as the term coupling
$\beta$ to the QD in the transformed Hamiltonian will vanish, as a
result of $\sin(\theta)=0$). In other words, in this case the QD
is spatially located in only one cavity and cannot interact with
the other, spatially distant and decoupled cavity. Fig.
\ref{phot_mol_theory} shows the numerically calculated cavity
transmission spectra (proportional to $\langle a^\dag
a\rangle+\langle b^\dag b\rangle$) when the QD is tuned across the
two cavity resonances. When the two cavities are coupled ($J\neq
0$), we observe anti-crossing between each cavity mode and the QD
(Fig. \ref{phot_mol_theory}a). However, only one anti-crossing is
observed when the cavities are not coupled (Fig.
\ref{phot_mol_theory}b).

The actual experiments are performed with self-assembled InAs QDs
embedded in GaAs, and the whole system is kept at cryogenic
temperatures ($\sim 10-25$ K) in a helium-flow cryostat. The
cavities used are linear three hole defect GaAs PC cavities
coupled via spatial proximity. The photonic crystal is fabricated
from a $160$ nm thick GaAs membrane, grown by molecular beam
epitaxy on top of a GaAs $(100)$ wafer. A low density layer of
InAs QDs is grown in the center of the membrane ($80$ nm beneath
the surface). The GaAs membrane sits on a $918$ nm sacrificial
layer of Al$_{0.8}$Ga$_{0.2}$As. Under the sacrificial layer, a
$10$-period distributed Bragg reflector, consisting of a
quarter-wave AlAs/GaAs stack, is used to increase collection into
the objective lens. The photonic crystal was fabricated using
electron beam lithography, dry plasma etching, and wet etching of
the sacrificial layer in diluted hydrofluoric acid, as described
previously \cite{article:eng07,article:majumdar09}.

\begin{figure}
\centering
\includegraphics[width=3.25in]{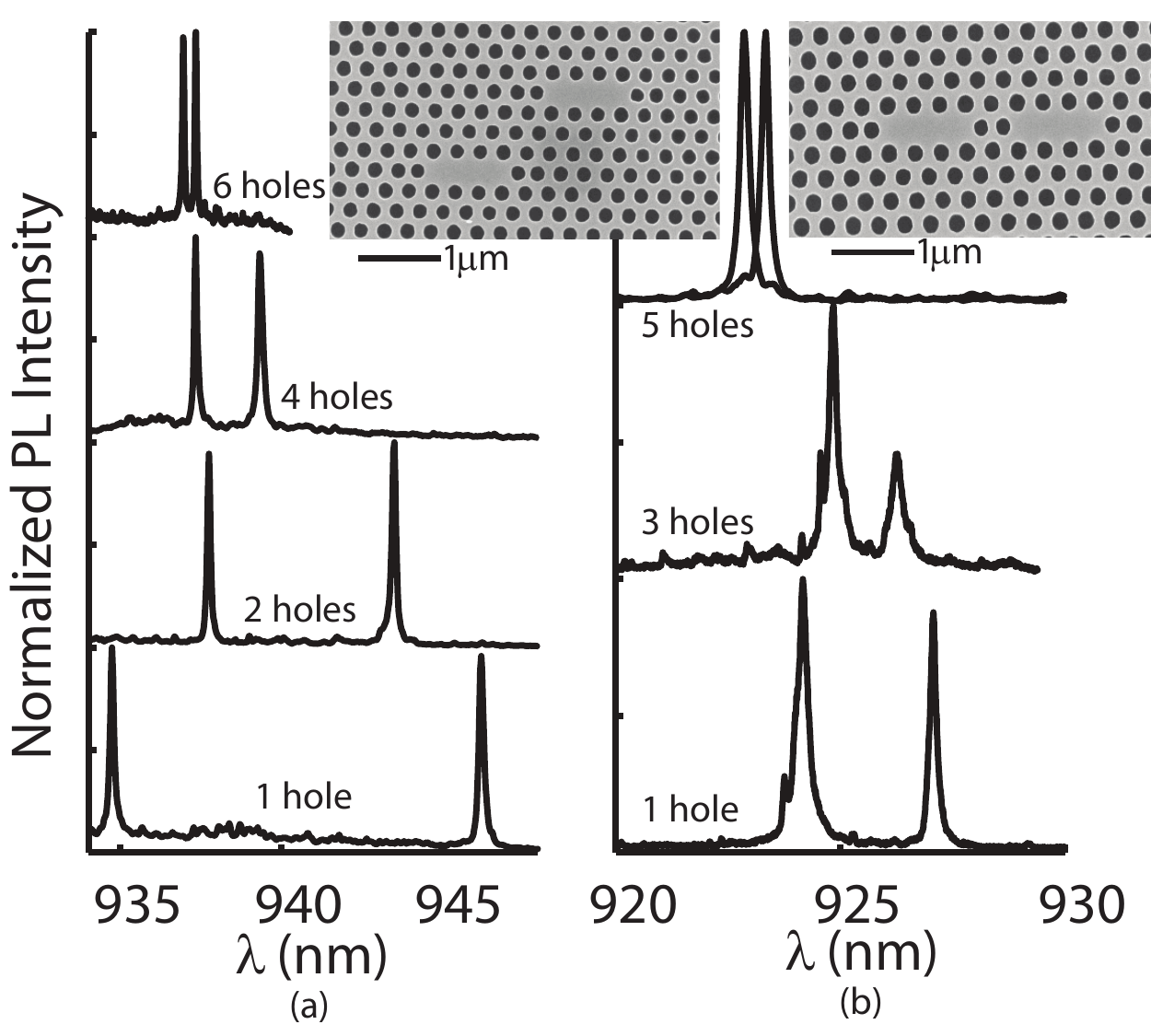}
\caption{Photoluminescence spectra of the coupled cavities for
different hole spacings between two cavities: (a) the cavities are
separated at an angle of $30^o$ (see the inset for a scanning
electron micrograph (SEM)); (b) the cavities are laterally
separated (see the inset for SEM). A decrease in the wavelength
separation between two cavity modes is observed with increasing
spatial separation between the cavities (i.e., with increasing
number of holes inserted in between the two cavities). A much
larger separation is observed in (a) when the cavities are coupled
at an angle compared to the lateral coupling (b).}
\label{SEM_separation}
\end{figure}

We fabricated two different types of coupled cavities: in one
case, the two cavities are offset at a $30^o$ angle (inset of Fig.
\ref{SEM_separation}a) and in the other the two cavities are
laterally coupled (inset of Fig. \ref{SEM_separation}b). In the
first case the coupling between the cavities is stronger as the
overlap between the electromagnetic fields confined in the
cavities is larger along the $30^o$ angle. Figs.
\ref{SEM_separation}a,b show the typical PL spectra of these two
different types of coupled cavities for different spacing between
the cavities. A clear decrease in the frequency separation between
the cavities is observed with increasing spatial separation. Note
that the consistency of this trend between different fabrication
runs already indicates that this frequency separation cannot be
purely due to the fabrication-related intrinsic detuning between
the two cavities. Nevertheless, it is very difficult to quantify
how much of the separation is due to coupling ($J$), and how much
is due to intrinsic detuning ($\Delta_o$) of the cavity
resonances. However, we will show that by observing the
anti-crossing between the QD and the two modes we can conclusively
determine both $J$ and $\Delta_o$.

\begin{figure}
\centering
\includegraphics[width=3.25in]{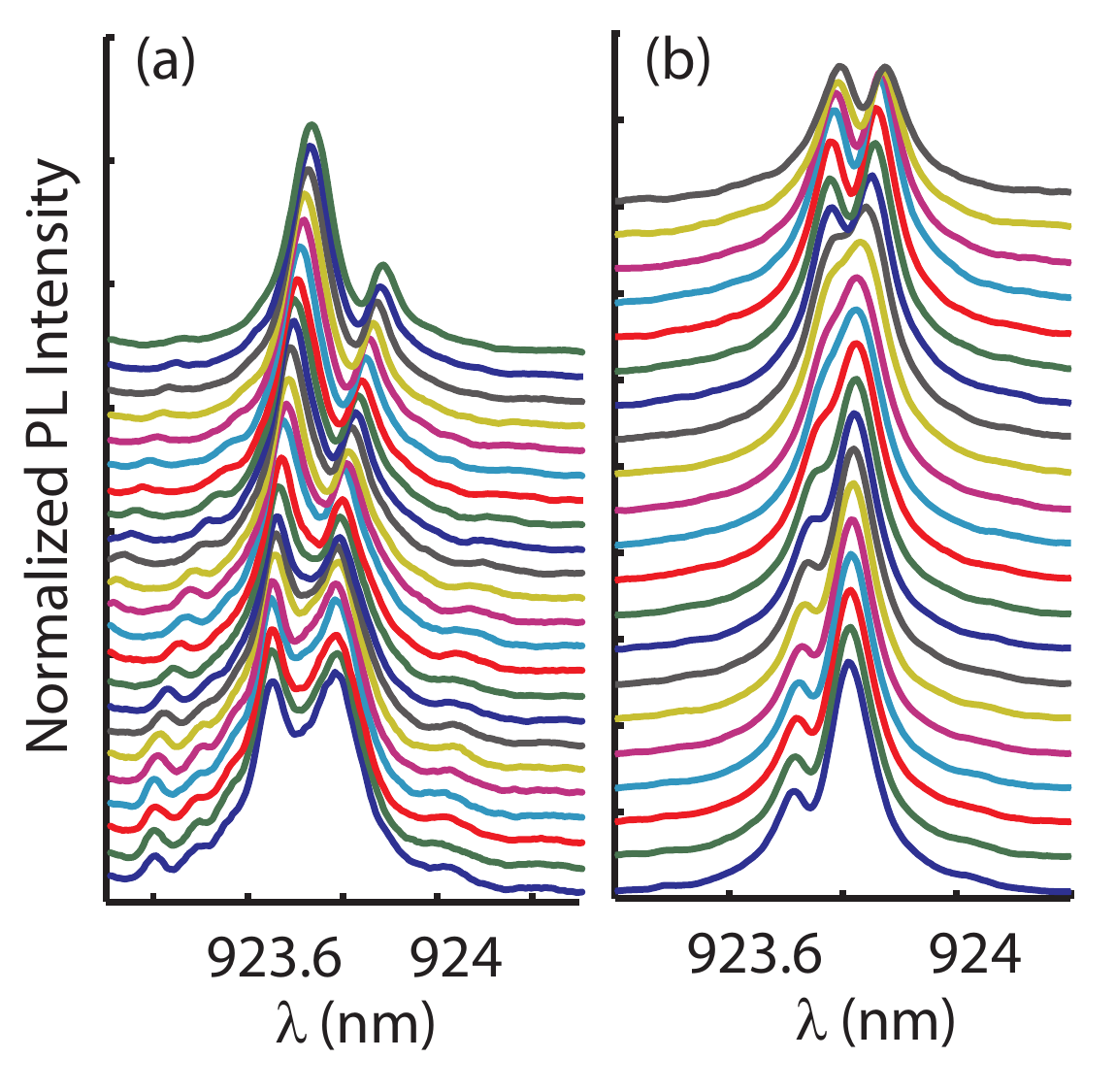}
\caption{(color online) Normalized PL intensity plotted when we
tune the QD across the cavity resonance by temperature: (a) before
nitrogen deposition (i.e., the QD is temperature tuned across the
longer wavelength resonance), and (b) after nitrogen deposition
(which red-shifts the cavity resonances and allows us to
temperature tune the QD across the shorter wavelength resonance).
Clear anti-crossings between the QD and the cavity are observed
for both super-modes. In both cases, the temperature is increased
from top to bottom (the plots are vertically offset for clarity).}
\label{Figure_AC_temp_tune}
\end{figure}

First, we investigate the strong coupling between a single QD and
the photonic molecule. For this particular experiment, we used a
photonic molecule consisting of cavities separated by $4$ holes
along the $30^o$ angle. In practice it is not trivial to tune the
QD over such a long wavelength range as required by the observed
separation of the two cavity peaks. Hence we use two different
tuning techniques: we tune the cavity modes by depositing nitrogen
on the cavity \cite{scherer_n2_tuning}, and then tune the QD
resonance across the cavity resonance by changing the temperature
of the system. We observe clear anti-crossings for both the modes
as shown in Figs. \ref{Figure_AC_temp_tune}a,b. Fig.
\ref{Figure_AC_temp_tune}a is obtained by temperature-tuning the
QD across the longer-wavelength cavity mode before nitrogen
deposition. We then perform the nitrogen deposition to red-shift
the cavity resonances, and repeat the temperature tuning. Fig.
\ref{Figure_AC_temp_tune}b shows the anti-crossing between the QD
and the shorter-wavelength cavity mode. The nitrogen and the
temperature tuning do not cause a significant change in the
coupling and the detuning between the cavities, as confirmed in
the experiments described below.

\begin{figure}
\centering
\includegraphics[width=3.25in]{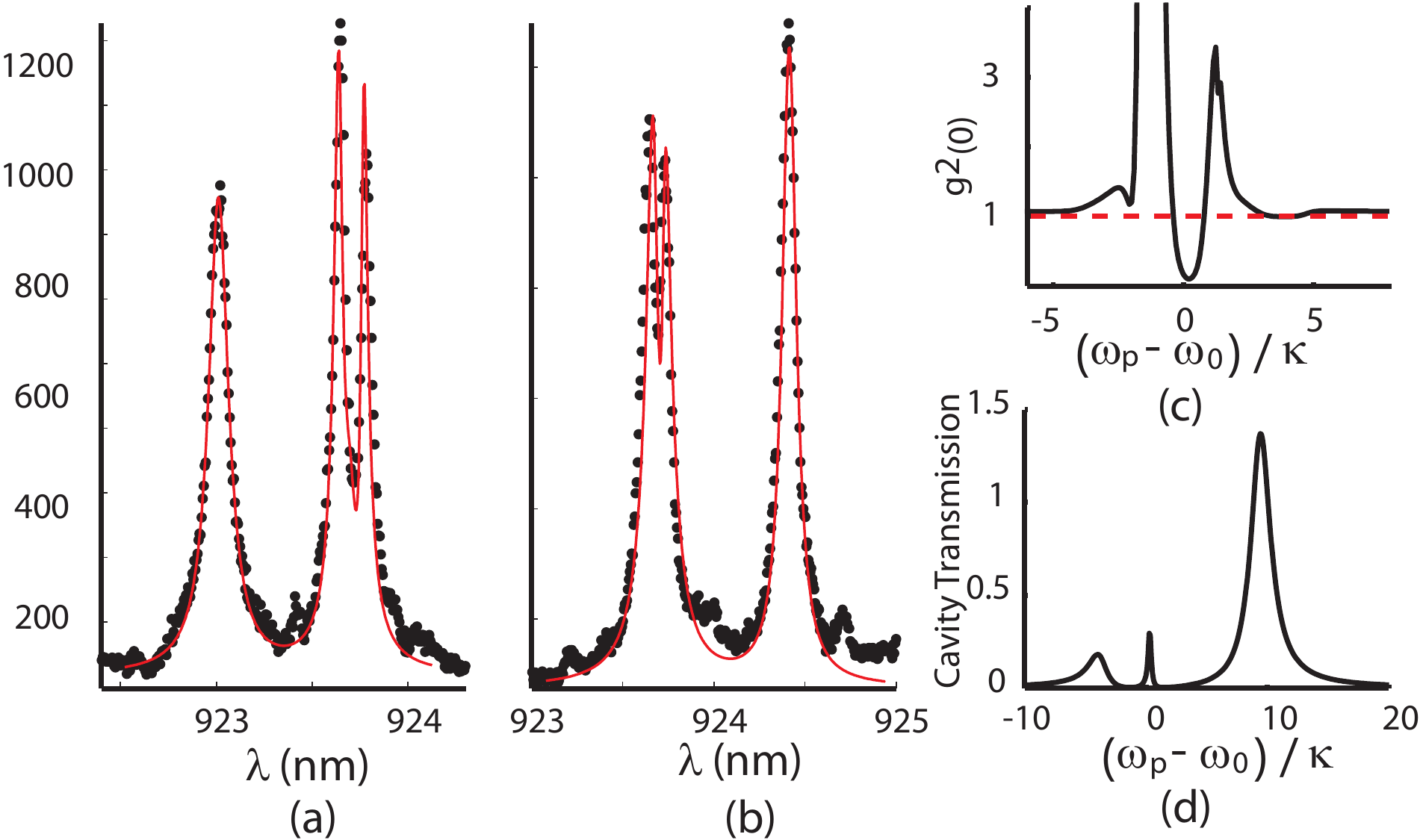}
\caption{(color online) QD-photonic molecule spectrum, (a) when
the QD is resonant with super-mode sm1 and (b) when the QD is
resonant with super-mode sm2. From the fit we extract the system
parameters (see text). Numerically simulated (c) second order
autocorrelation $g^2(0)$ and (d) transmission from cavity $b$, as
a function of laser frequency, with the experimental system
parameters that were extracted from the fits.}
\label{fig_fit_data}
\end{figure}

We perform curve-fitting for the PL spectra when the QD is
resonant to the cavity super-modes and estimate the system
parameters (Figs. \ref{fig_fit_data}a,b). The super-mode at
shorter (longer) wavelength is denoted as sm1 (sm2). As the
detuning between the super-modes is much larger than the vacuum
Rabi spitting caused by the QD, we can assume that when the QD is
resonant to sm1(2), its interaction with sm2(1) is negligible.
Therefore, we can fit the PL spectra of sm1 (sm2) modes exhibiting
Rabi splitting individually. For sm1, we extract from the fit the
field decay rate $\kappa_1/2\pi=16.7$ GHz and the QD-field
interaction strength $g_1/2\pi=23.7$ GHz (Fig.
\ref{fig_fit_data}a); for sm2, $\kappa_2/2\pi=22.4$ GHz and
$g_2/2\pi=14.2$ GHz (Fig. \ref{fig_fit_data}b). We note that we
can achieve very high quality factors ($\sim 7,000-10,000$) of the
coupled cavity modes as seen from the extracted $\kappa$ values.
We also estimate the total detuning between two observed modes as
$\Delta/2\pi=0.7$ and $0.72$ GHz before and after nitrogen tuning.
This minimal difference in $\Delta$ resulting from the nitrogen
tuning does not impact our further analysis, and we take $\Delta$
to be the average of these two values. The change in the cavity
field decay rates arising from the nitrogen deposition is also
minimal. From these data, we use the relations
$\theta=\arctan(g_2/g_1)$, $\tan(2\theta)=-2J/\Delta_o$ and
$\Delta=\sqrt{4J^2+\Delta_o^2}$ to obtain: $J/2\pi \approx 110$
GHz and $\Delta_o/2\pi \approx 118$ GHz.

We now numerically simulate the performance of such a QD-photonic
molecule for generation of sub-Poissonian light using the quantum
optical master equation approach \cite{majumdar_phonon_11}. Two
bare cavity modes are separated by $\Delta_o/2\pi=118$ GHz; a QD
is resonant and strongly coupled to one of the modes ($a$) with
interaction strength $g/2\pi=27.6$ GHz ($g=\sqrt{g_1^2+g_2^2}$,
where $g_1$ and $g_2$ are the two values of QD-cavity interaction
strengths obtained by fitting the PL spectra); mode $b$ is the
empty cavity. The mode $b$ is driven and the second order
autocorrelation $g^2(0)=\frac{\langle b^\dag b^\dag b
b\rangle}{\langle b^\dag b \rangle^2}$ of the transmitted light
through cavity $b$ is calculated \cite{imma_ph_molecule}. We also
assume the two cavities to have the same cavity decay rate, which
is an average of the cavity decay rates measured from the two
super-modes. Note however that, having slightly different decay
rates does not significantly affect the performance of the system.
The numerically simulated cavity $b$ transmission and $g^2(0)$ of
the transmitted light is shown in Figs. \ref{fig_fit_data}c,d. We
note that with our system parameters we can achieve strongly
sub-Poissonian light with $g^2(0)\sim 0.03$. Unfortunately, in
practice it is very difficult to drive only one cavity mode
without affecting the other mode due to the spatial proximity of
two cavities. This individual addressability is critical for good
performance of the system \cite{imma_ph_molecule} and to retain
such a capability in a photonic molecule the cavities should be
coupled via a waveguide \cite{NodaWgCoupling}.

\begin{figure}
\centering
\includegraphics[width=3.25in]{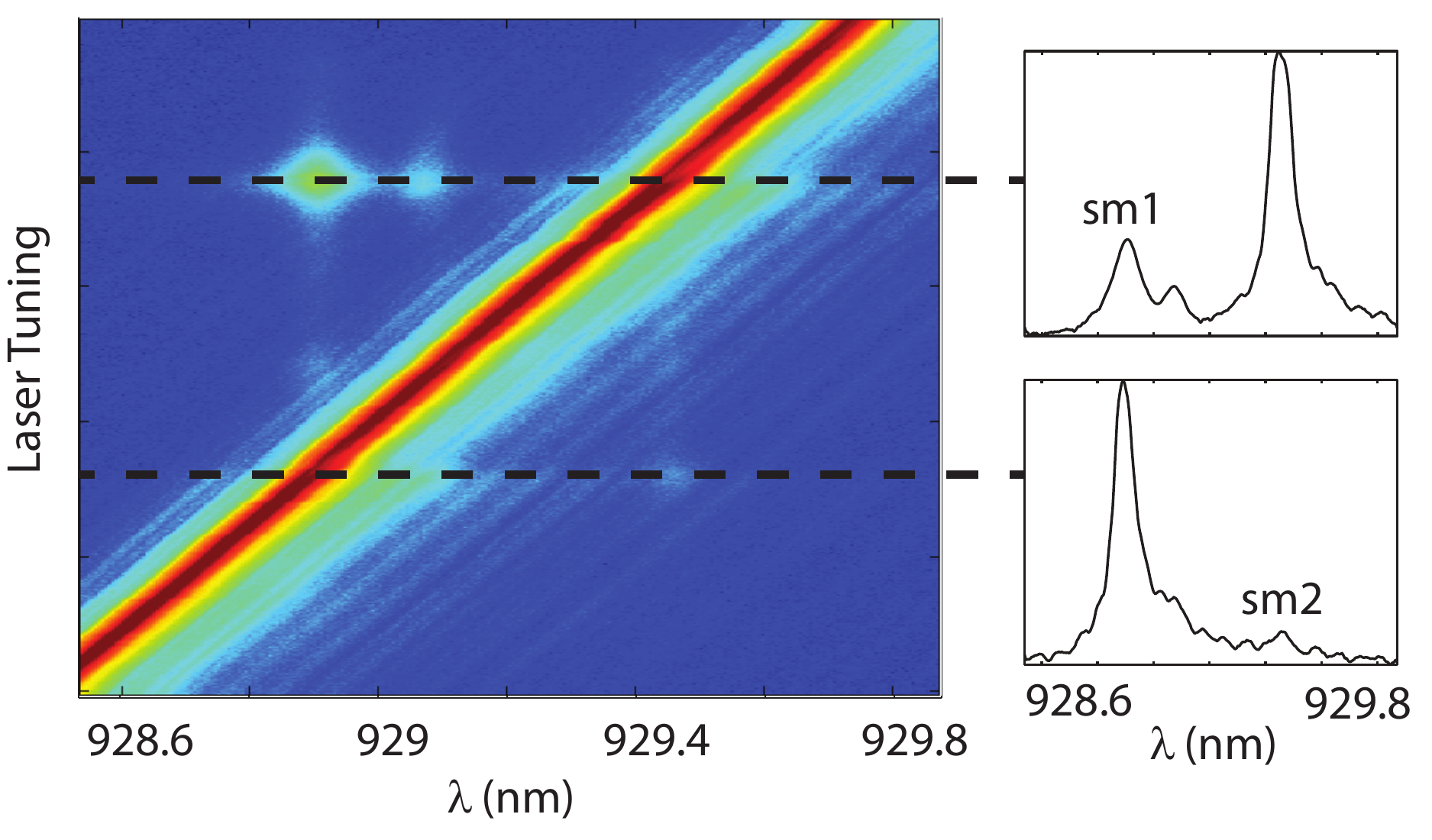}
\caption{(color online) Off-resonant interaction between two
coupled cavities and a QD. We scan the laser across both coupled
modes, and observe emission from the off-resonant super-mode,
under excitation of the other super-mode. A close-up spectrum for
each resonance shows the relative position of the laser and the
cavity modes.} \label{figure_nrdc}
\end{figure}

Finally, as a further demonstration of cQED effects in this
system, we report off-resonant interaction between the coupled
cavities and the QD, similar to the observations in a single
linear three hole defect cavity \cite{article:majumdar10} and a
nano-beam cavity \cite{armand_APL}. This experiment was performed
on a different QD-photonic molecule system than the one where we
observed strong coupling. Fig. \ref{figure_nrdc} shows the spectra
indicating off-resonant coupling between the cavities and the QD.
Under resonant excitation of the supermode at longer wavelength
(sm2), we see pronounced emission from both sm1 and a nearby QD.
Similarly, under resonant excitation of sm1, we see emission from
sm2, although the emission is much weaker. We exclude the presence
of any nonlinear optical processes by performing a laser-power
dependent study of the cavity emission, which shows a linear
dependence of the cavity emission on the laser power (not shown
here).

In summary, we demonstrated strong coupling of a single QD to a
photonic molecule in a photonic crystal platform. Clear
anti-crossings between the QD and both super-modes of the photonic
molecule were observed, showing conclusive evidence of
inter-cavity coupling. From the anti-crossing data we were able to
separate the contributions of the inter-cavity coupling and
intrinsic detuning to the cavity mode splitting. We have also
reported observation of off-resonant cavity-cavity and cavity-QD
interaction in this type of system. Such a system could be
employed for non-classical light generation (as theoretically
studied in this article), and represents a building block for an
integrated nanophotonic network in a solid-state cQED platform.

The authors acknowledge financial support provided by the Office
of Naval Research (PECASE Award; No: N00014-08-1-0561), DARPA
(Award No: N66001-12-1-4011), NSF (DMR-0757112) and Army Research
Office (W911NF-08-1-0399). A.R. is also supported by a Stanford
Graduate Fellowship. We acknowledge Dr. Hyochul Kim and Dr. Pierre
Petroff for providing the quantum dot sample. This work was
performed in part at the Stanford Nanofabrication Facility of
NNIN, supported by the National Science Foundation.
\bibliography{Phot_mol_CQED}
\end{document}